\documentclass[aps,english,preprint,nofootinbib]{revtex4}

\usepackage{hyperref}
\usepackage{graphicx}
\usepackage{amsmath, amssymb}
\usepackage{babel}
\usepackage{color}
\usepackage{mathrsfs}
\usepackage{slashed}

\usepackage[T1]{fontenc} 
\usepackage{ulem}
\usepackage{float}
\usepackage{epstopdf}
\usepackage{subfigure}
\usepackage{bbm}
\usepackage{makecell}

\begin{document}

\title{The dependence of overlap topological charge density on Wilson mass parameter}

\author{Zhen Cheng }
\author{Jian-bo Zhang}

\affiliation{Department of Physics, Zhejiang University, Zhejiang 310027, P.R. China}


\begin{abstract}
In this paper, we analyze the dependence of the topological charge density from the overlap operator on the Wilson mass parameter in the overlap kernel by the symmetric multi-probing source (SMP) method. We observe that the non-trivial topological objects are removed as the Wilson mass is increased. A comparison of topological charge density calculated by the SMP method using fermionic definition with that of gluonic definition by the Wilson flow method is shown. A matching procedure for these two methods is used. We find that there is a best match for topological charge density between gluonic definition with varied Wilson flow time and fermionic definition with different Wilson mass. By using the matching procedure, the proper flow time of Wilson flow in the calculation of topological charge density can be estimated. As the lattice spacing $a$ decreases, the proper flow time also decreases, as expected. 
\end{abstract}

\maketitle

\section{Introduction}
Topological charge $Q$ and its density $q\left(x\right)$ play an important role in the study of the non-trivial topological
structure of QCD  vacuum. Topological properties have important
phenomenological implications, such as $\theta$ dependence and spontaneous
chiral symmetry breaking. The confinement may also be related to nontrivial
topological properties \cite{Schierholz:1994pb,Witten:1978bc,Diakonov:1995ea}. The topology of QCD gauge fields is a non-perturbative
issue, therefore, lattice method is a good choice to investigate it
from first principles. Lattice QCD is powerful for studying the
topological structure of the vacuum. There are many definitions
of the topological charge for a lattice gauge field \cite{Cichy:2014qta,Muller-Preussker:2015daa,Alexandrou:2017hqw}. These definitions can be characterized either as gluonic or
fermionic. In the fermionic definition, topological charge $Q$ is the number of
zero modes of the Dirac operator \cite{Atiyah:1971rm,Hasenfratz:1998ri}. On the other hand,
topological charge can be given by the field strength tensor (gluonic definition)
on the lattice, and this definition approaches fermionic definition
in the limit $a\rightarrow0$ \cite{Belavin:1975fg,Fujikawa:1998if,Kikukawa:1998pd}.

The overlap Dirac operator is a solution of the Ginsparg-Wilson equation \cite{Neuberger:1997fp,Neuberger:1998wv}, and the topological charge defined from the overlap fermion will be an exact integer. In the traditional method the point source is used in the calculation of topological charge density \cite{Horvath:2003yj,Ilgenfritz:2007xu}, which makes the computation on the large lattice almost impossible. In order to reduce the computational cost, the symmetric source (SMP) method is introduced to calculate the topological charge density \cite{Xiong:2019pmh}. As the Wilson mass parameter $m$ varies, the value of $Q$ may change \cite{Narayanan1995,Edwards:1998sh,Narayanan:1997sa,Zhang:2001fk}. The topological charge density $q\left(x\right)$ has strong correlation with low-lying modes of the Dirac operator, which strongly influences how quarks propagate through the vacuum. So the topological charge density $q\left(x\right)$ is a useful probe of the gauge field. We visualize the topological charge density and view the detailed extra information \cite{Vege:2019nee}. On the other hand, the topological charge can not show the details of the QCD vacuum. So we will focus on the topological charge density $q\left(x\right)$ in the study of the topological properties of the QCD vacuum. We will show an analysis of the topological charge $q\left(x\right)$, obtained using fermionic definition with different values of $m$ and gluonic definition with different Wilson flow time. Unlike in the case of Ref. \cite{Moran:2010rn}, which studied just one time slice, we consider all time slices and show more details on the topological charge density with different topological charge. By analyzing the topological charge density $q\left(x\right)$, we can obtain a great amount of information about the underlying topological structure. We show a comparison with the gluonic topological charge density that is calculated after the application of the Wilson flow method. This comparison will be shown by calculating the matching parameter $\Xi_{AB}$, which will be defined later. $\Xi_{AB}$ will be used to measure the match of the topological charge density between the fermionic definition and gluonic definition. The best match is found when the matching parameter is nearest to $1$. The proper flow time of Wilson flow in the gluonic definition can also be obtained by analysing the matching procedure. The behavior of the proper flow time toward the continuum limit is also discussed. 

\section{Simulation details}
The pure gauge lattice configurations were generated using a tadpole improved, plaquette plus
rectangle gauge action through pseudo-heat-bath algorithm \cite{Luscher:1984xn,Bonnet:2001rc}.
This gauge action at tree-level $\mathcal{O}\left(a^{2}\right)$-improved
is defined as
\begin{equation}
S_{G}= \frac{5\beta}{3}\sum_{x\mu\nu\atop \nu>\mu}\operatorname{Re}\operatorname{Tr}\left[1-P_{\mu\nu}(x)\right]-\frac{\beta}{12u_{0}^{2}}\sum_{{x\mu\nu\atop \nu>\mu}}\operatorname{Re}\operatorname{Tr}\left[1-R_{\mu\nu}(x)\right],
\end{equation}
where $P_{\mu\nu}$ is the plaquette term. The link product $R_{\mu\nu}\left(x\right)$
denotes the rectangular $1\times2$ and $2\times1$ plaquettes. The mean link, $u_{0}$
is the tadpole improvement factor that largely corrects for the quantum renormalization of coefficient
for the rectangles relative to the plaquette. $u_{0}$ is given by
\begin{equation}
u_{0}=\left(\frac{1}{3}\text{Re}\thinspace\text{Tr}\left\langle P_{\mu\nu}\left(x\right)\right\rangle \right)^{1/4}.
\end{equation}

In the fermionic definitions, we use the overlap operator to calculate
topological charge density. The massless overlap Dirac operator
is given by \cite{Zhang:2001fk}
\begin{align}
D_{\text{ov}} & =\left(\mathbbm{1}+\frac{D_{\rm{W}}}{\sqrt{D_{\rm{W}}^{\dagger}D_{\rm{W}}}}\right),
\end{align}
where $D_{\rm{W}}$ is the Wilson Dirac operator,
\begin{equation}
D_{{\rm W}}  =\delta_{a,b}\delta_{\alpha,\beta}\delta_{i,j}-\kappa\sum_{\mu=1}^{4}[\left(\mathbbm{1}-\gamma_{\mu}\right)_{\alpha\beta} U_{\mu}\left(i\right)_{ab}\delta_{i,j-\hat{\mu}}+\left(\mathbbm{1}+\gamma_{\mu}\right)_{\alpha\beta}U_{\mu}^{\dagger}\left(i-\hat{\mu}\right)_{ab}\delta_{i,j+\hat{\mu}}],
\end{equation}
and $\kappa$ is the hopping parameter,
\begin{equation}
\kappa=\frac{1}{2\left(-m+4\right)}. \label{hop-par}
\end{equation}
In the overlap formalism $\kappa$ has to be in the range $\left(\kappa_c,0.25\right)$ for $D_{\rm{ov}}$ to describe a single massless Dirac fermion, and $\kappa_c$ is the critical value of $\kappa$ at which the
pion mass extrapolates to zero in the simulation with ordinary Wilson fermions. We call $m$ in Eq.(\ref{hop-par}) as the Wilson mass parameter. In this work, we choose parameter $\kappa$ as the input parameter.

The overlap topological charge density can be calculated as follows,
\begin{equation}
q_{\text{ov}}\left(x\right)=\frac{1}{2}\text{Tr}_{c,d}\left(\gamma_{5}D_{\text{ov}}\left(x\right)\right)=\text{Tr}_{c,d}\left(\tilde{D}_{\text{ov}}\left(x\right)\right), \label{q-ov}
\end{equation}
where the trace is over the color and Dirac indices. It is well known that the traditional way of computing  $q_{\text{ov}}\left(x\right)$ with point source is almost impossible for the large lattice volume.

In order to avoid the high computational effort in the calculation
of the $q\left(x\right)$ with point source, we apply the symmetric multi-probing (SMP) method to calculate $q\left(x\right)$ \cite{Xiong:2019pmh,Cheng:2020dlb},

\begin{align}
q_{{\rm smp}}\left(x\right) & =\sum_{\alpha,a}\psi\left(x,\alpha,a\right)\left(\tilde{D}_{{\rm ov}}\left(x\right)\right)\phi_{P}\left(S\left(x,P\right),\alpha,a\right) \nonumber \\
& =\sum_{\alpha,a}\psi\left(x,\alpha,a\right)\left(\tilde{D}_{{\rm ov}}\left(x\right)\right)\psi\left(x,\alpha,a\right)  \nonumber \\
& \thinspace\thinspace\thinspace\thinspace+\sum_{y\in S\left(x,P\right)}^{y\ne x}\psi\left(x,\alpha,a\right)\left(\tilde{D}_{{\rm ov}}\left(x\right)\right)\psi\left(y,\alpha,a\right)  \nonumber \\
& \approx\sum_{\alpha,a}\psi\left(x,\alpha,a\right)\left(\tilde{D}_{{\rm ov}}\left(x\right)\right)\psi\left(x,\alpha,a\right), \label{q-smp}
\end{align}
where $x$ is the seed site at site $\left(x_{1},x_{2},x_{3},x_{4}\right)$ and $y$ are the other lattice
sites belonging to the set $S\left(x,P\right)$. $\phi_{P}\left(S\left(x,P\right),\alpha,a\right)$
is the SMP source vector, and $\psi$ is the normalized point source
vector. $S\left(x,P\right)$ represents the sites with the same color
of $x$ obtained by the symmetric coloring scheme $P\left(\frac{n_{s}}{d},\frac{n_{s}}{d},\frac{n_{s}}{d},\frac{n_{t}}{d},\text{mode}\right)$.
$n_{s}$ and $n_{t}$ are the spatial and temporal sizes of
the lattice, $d$ is the minimal distance of the coloring scheme and $\text{mode}=0, 1, 2$ corresponds
to the Normal, Split and Combined $\text{mode}$ for scheme $P$, and the number of SMP sources that cover all lattice site are $12d^{4}$, $24d^{4}$ and $6d^{4}$, respectively. The term in the third line of Eq.(\ref{q-smp}) is the summation of space-time off-diagonal elements of $\tilde{D}_{{\rm ov}}\left(x\right)$. Because of the space-time locality of $D_{\rm ov}$, this term can be regarded as the error in the calculation of topological charge density. If we choose the proper scheme $P$ of the SMP source, we can neglect the error term and get the last line in the Eq.(\ref{q-smp}).

However, the number of the normalized point source is $12N_{L}$, and $N_L=N_{x}N_{y}N_{z}N_{t}$ is the lattice volume. It shows that SMP method is much more cheaper than the point source method in the calculation of topological charge density with fermionic definition, especially for large lattice volume. The topological charge by using SMP method is denoted as $Q_{\rm smp}$, given by
\begin{equation}
Q_{\rm smp}=\sum_{x}q_{{\rm smp}}\left(x\right).
\end{equation}

Gradient flow is a non-perturbative smoothing procedure, which has
been proven to have well-defined numerical and perturbative properties. The gradient flow is defined as the solution of the evolution equations \cite{Luscher:2010iy,Bonati:2014tqa,Alexandrou:2017hqw,Alexandrou:2015yba}
\begin{equation}
\dot{V}_{\mu}\left(x,\tau\right)=-g_{0}^{2}\left[\partial_{x,\mu}S_{G}\left(V\left(\tau\right)\right)\right]V_{\mu}\left(x,\tau\right),\thinspace\thinspace\thinspace\thinspace V_{\mu}\left(x,0\right)=U_{\mu}\left(x\right),\label{eq:gra-flow}
\end{equation}
where $\tau$ is the dimensionless gradient flow time (Wilson flow time in this work) and $g_{0}^{2}\partial_{x,\mu}S_{G}\left(U\right)$ is given by
\begin{align}
g_{0}^{2}\partial_{x,\mu}S_{G}\left(U\right) & =2i\sum_{a}T^{a}\text{Im}\thinspace\text{Tr}\left[T^{a}\Omega_{\mu}\right]\nonumber \\
& =\frac{1}{2}\left(\Omega_{\mu}\left(x\right)-\Omega_{\mu}^{\dagger}\right)-\frac{1}{6}\text{Tr}\left(\Omega_{\mu}\left(x\right)-\Omega_{\mu}^{\dagger}\right),
\end{align}
where $T^{a}\thinspace(a=1,2,\cdots,8)$ are the Hermitian generators of the $\rm{SU}(3)$ group. $\Omega_{\mu}=U_{\mu}\left(x\right)X_{\mu}^{\dagger}\left(x\right)$, and $X_{\mu}\left(x\right)$ is the so-called staples.

In practice, the gradient flow moves the gauge configuration along the
steepest descent direction in the configuration space, such as along
the gradient of the action. The chosen sign in the evolution equations
leads to a minimization of the action, which is as expected. We use
the third order Runge-Kutta method to obtain the solution of the flow
Eq. (\ref{eq:gra-flow}). The gluonic definition of topological charge density in Euclidean spacetime is defined as
\begin{equation}
q\left(x\right)=\frac{1}{32\pi^{2}}\epsilon_{\mu\nu\rho\sigma}\text{Tr}\left[F_{\mu\nu}F_{\rho\sigma}\right],
\end{equation}
with $F_{\mu\nu}$ the gluonic field strength tensor. The topological charge of a gauge field is the four-dimensional integral over space-time of the topological charge density,
\begin{equation}
Q=\int{\rm d}^{4}x\thinspace q\left(x\right).
\end{equation}
The most common definition of the topological charge density in lattice discretization is the clover definition given by
	\begin{equation}
	q_{L}^{\text{clov}}\left(x\right)=\frac{1}{32\pi^{2}}\epsilon_{\mu\nu\rho\sigma}{\rm Tr}\left[C_{\mu\nu}^{\text{clov}}C_{\rho\sigma}^{\text{clov}}\right],
	\end{equation}
where $C_{\mu\nu}^{\text{clov}}$ is the usual clover leaf.

The field strength tensor $F_{\mu\nu}$ used in this work is a 3-loop $\mathcal{O}\left(a^{4}\right)$-improved and defined as \cite{BilsonThompson:2002jk},
\begin{equation}
F_{\mu\nu}^{\text{Imp}}=\frac{27}{18}C^{\left(1,1\right)}-\frac{27}{180}C^{\left(2,2\right)}+\frac{1}{90}C^{\left(3,3\right)}, \label{q-impro}
\end{equation}
and $C^{\left(m,n\right)}$ denotes the three $m\times n$ loops used to construct the clover term, and $C^{\left(1,1\right)}$ is the clover leaf mentioned above.

The $q\left(x\right)$ calculated after Wilson flow to the gauge configuration is denoted as $q_{\text{wf}}\left(x\right)$. The topological charge obtained by Wilson flow is denoted as $Q_{\rm wf}$, given by
\begin{equation}
Q_{{\rm wf}}=\sum_{x}q_{{\rm wf}}\left(x\right).
\end{equation} 

In order to fairly compare the two definitions for the topological
charge density with the varied Wilson mass parameter, we will calculate the matching parameter $\Xi_{AB}$, given by \cite{Bruckmann:2006wf}
\begin{equation}
\Xi_{AB}=\frac{\chi_{AB}^{2}}{\chi_{AA}\chi_{BB}},
\end{equation}
with
\begin{equation}
\chi_{AB}=\frac{1}{V}\sum_{x}\left(q_{A}\left(x\right)-\bar{q}_{A}\right)\left(q_{B}\left(x\right)-\bar{q}_{B}\right),
\end{equation}
where $\bar{q}$ denotes the mean value of $q\left(x\right)$, and
in this work $q_{A}\left(x\right)\equiv q_{\text{smp}}\left(x\right)$,
$q_{B}\left(x\right)\equiv q_{\text{wf}}\left(x\right)$. When the
$\Xi_{AB}$ is nearest to $1$, the best match is found \cite{Moran:2010rn}. When the best match is reached, the flow time $\tau$ is called the proper flow time of Wilson flow, denoted as $\tau_{\rm pr}$. In this work, the step length for numerical intergration of Wilson flow is $\delta \tau=0.005$.

We also calculate the factor $Z_{{\rm calc}}$, defined as
\begin{equation}
Z_{{\rm calc}}\equiv\frac{\sum_{x}\left|q_{{\rm ov}}\left(x\right)\right|}{\sum_{x}\left|q_{{\rm wf}}\left(x\right)\right|}.
\end{equation}
Because the topological charge of gluonic definition is not always an integer, $Z_{\rm calc}$ is needed in the visualization of the matching procedure. The matching parameter $\Xi_{AB}$ is independent of the value of $Z_{\rm calc}$.

We will analyse $q\left(x\right)$ of all time slices on lattices of $16^{4}$, $24^{3}\times48$ and $32^{4}$ at the inverse coupling, $\beta=4.50$, $\beta=4.80$ and $\beta=5.0$, corresponding to the lattice spacing $a=0.1289$, $0.0845$ and $0.0655\thinspace\text{fm}$  respectively. In this work, we calculate topological charge density of two configurations for lattice volumes $16^{4}$ and $24^{4}\times48$, and one configuration for $32^{4}$. We show the visualization of the topological charge density and apply the matching procedure to obtain the proper flow time of Wilson flow in the calculation of topological charge density.

\section{$\boldsymbol{q\left(x\right)}$ for different methods \label{matching-procedure}}
\begin{figure}[H]
	\centering
	\includegraphics[width=3.0in]{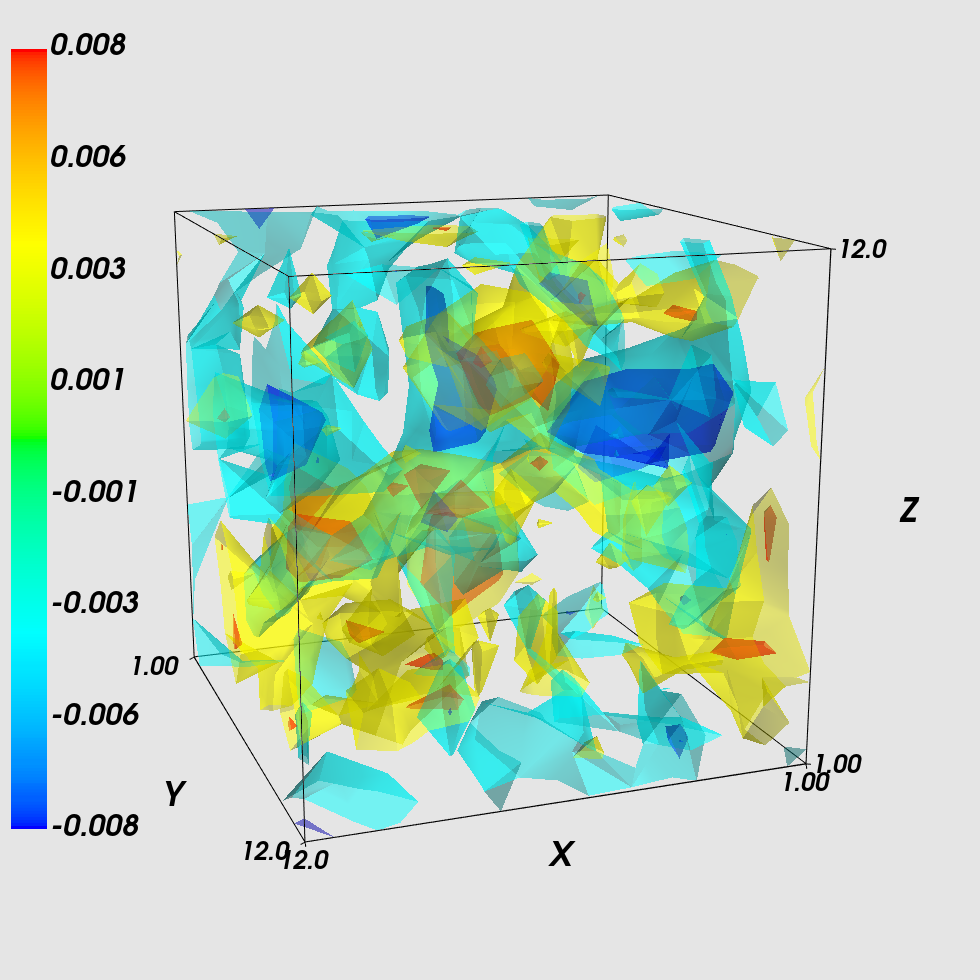}\,\thinspace\includegraphics[width=3.0in]{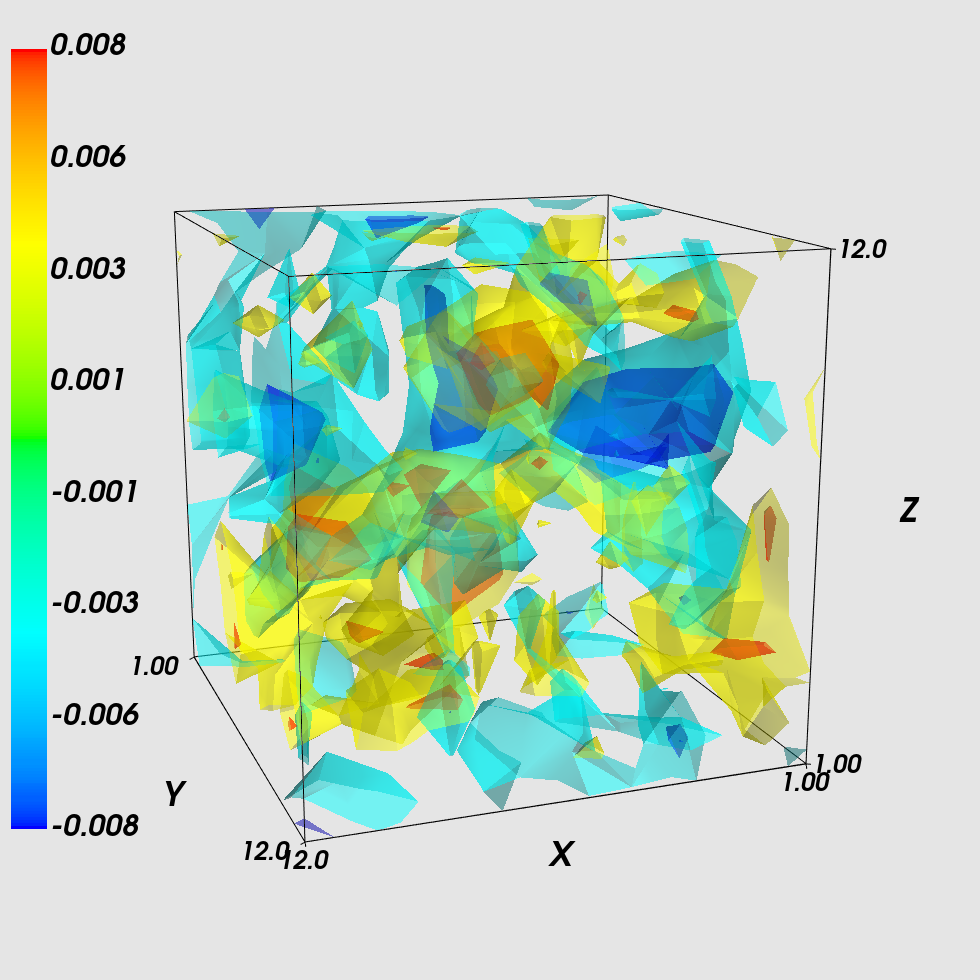}	
	\caption{$q_{{\rm ps}}\left(x\right)$ and $q_{{\rm smp}}\left(x\right)$ on lattice $12^{3}\times24$ by the
		point source and SMP source for the $t=12$ slice, and other time slices are similar. To the naked eyes, they are almost the same. The parameters, $\kappa=0.21$ and $\beta=4.80$, are the same for both setups. In the figures, we use a color cutoff method shown as the color map.  Left: Point source,
		Right: SMP source. \label{fig:q_sp-q_smp}}
\end{figure}

Before we proceed the analysis, we first demonstrate that the SMP method with proper $d$ is a good method to evaluate the topological charge density with overlap Dirac operator. When the source in the calculation of topological charge is point source, $q_{{\rm ps}}\left(x\right)$ is an exact result of $q_{{\rm ov}}\left(x\right)$. It is reasonable to use $q_{{\rm ps}}\left(x\right)$ as a benchmark for comparison. Due to the high computational cost, we only make a comparison of the point source and the SMP source in eq. (\ref{q-ov}) on $12^{3}\times24$  lattice with $\beta=4.8$,  $\kappa=0.21$. We show $q_{{\rm ps}}\left(x\right)$ obtained using the point source method and $q_{{\rm smp}}\left(x\right)$ using the SMP method with $d=6$ in Fig. \ref{fig:q_sp-q_smp}. In order to show visualization more clear, we use a cutoff method, shown as the color map. The same cuttoff procedure is used in other visualized figures.  It shows that $q_{{\rm smp}}\left(x\right)$ is highly matched with $q_{{\rm ps}}\left(x\right)$. The matching parameter $\Xi_{{\rm AB}}$ for these two methods is $0.9997$, and we can barely see the difference by naked eyes. It shows that error caused by the space-time off-diagonal elements of $D_{\rm ov}$ is indeed very small when the distance parameter $d$ of SMP source is proper. Thus the SMP method is a good choice to calculate the topological charge density while the parameter $d$ is large enough. It is expected that a better match goes with a larger distance $d$ \cite{Xiong:2019pmh}.

In order to obtain more precise topological charge density, we choose the distance parameter $d=8$ in the SMP method on $16^{4}$, $24^{3}\times48$, $32^{4}$ ensembles. The results are summarized in Table \ref{tab:table-x16t16-conf1} $\sim$ \ref{tab:table-x32t32-conf1}.  We only show the visualization of topological charge density $q\left(x\right)$ for lattice volume $24^{3}\times48$ at $\beta=4.80$ as an example. Results for other lattice ensembles are similar. In these calculations of $q\left(x\right)$ with $d=8$ by the SMP method, we find that the topological charge $Q_{\rm smp}$ is very close to an integer and it is not always the same for the same ensemble with different $\kappa$ values. This is acceptable as zero crossings in the spectral flow of the $\tilde{D}_{\text{ov}}\left(x\right)$ occur for different $\kappa$ on lattice \cite{Edwards:1998sh,Alexandrou:2017hqw}. From the tables, it also shows that $Q_{\rm smp}$ is very sensitive to $\kappa$ on the coarsest lattice. $Q_{\rm smp}$ changes a little on the finer lattice, and is stable versus the change of $\kappa$ on the finest lattice. This sensitivity is probably the effect of finite lattice space $a$. All results indicates that even though topological charge $Q_{\rm smp}$ from overlap fermions is always an integer, its value is not unique, and it depends on Wilson mass parameter $m$. 

\begin{figure}
	\centering
	\subfigure{
		\begin{minipage}[t]{0.36\textwidth}
			\includegraphics[width=2.25in]{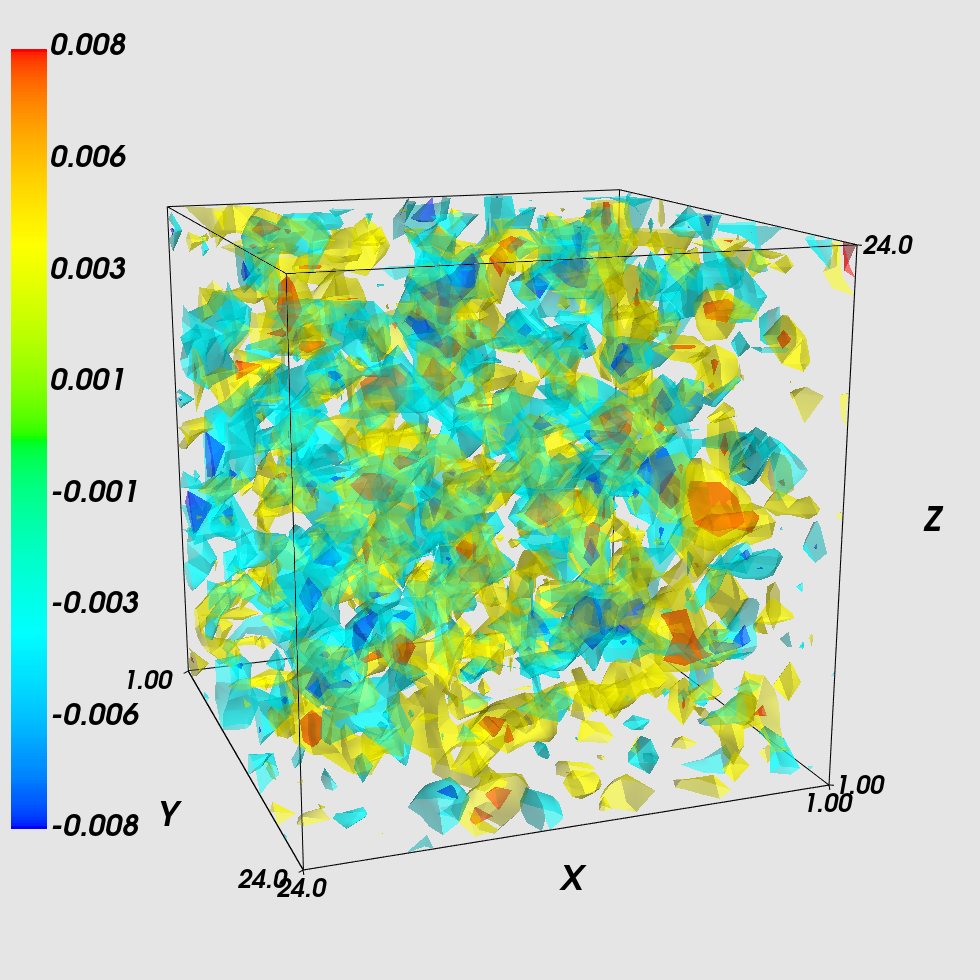}\\
			\centerline{$\kappa=0.23$}
		\end{minipage}
		\begin{minipage}[t]{0.36\textwidth}
			\includegraphics[width=2.25in]{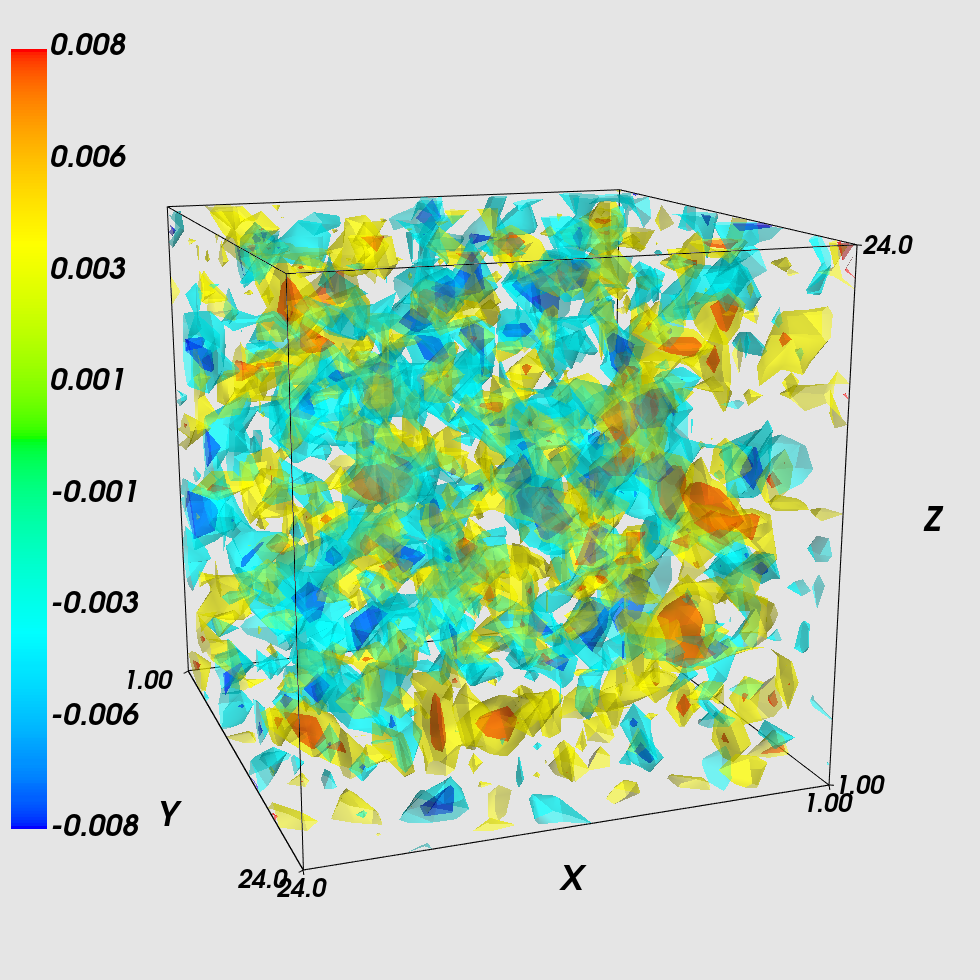}\\
			\centerline{$\tau_{{\rm pr}}=0.295$}			
	\end{minipage}	}
	\hspace{-15mm}
	\subfigure{
		\begin{minipage}[t]{0.36\textwidth}
			\includegraphics[width=2.25in]{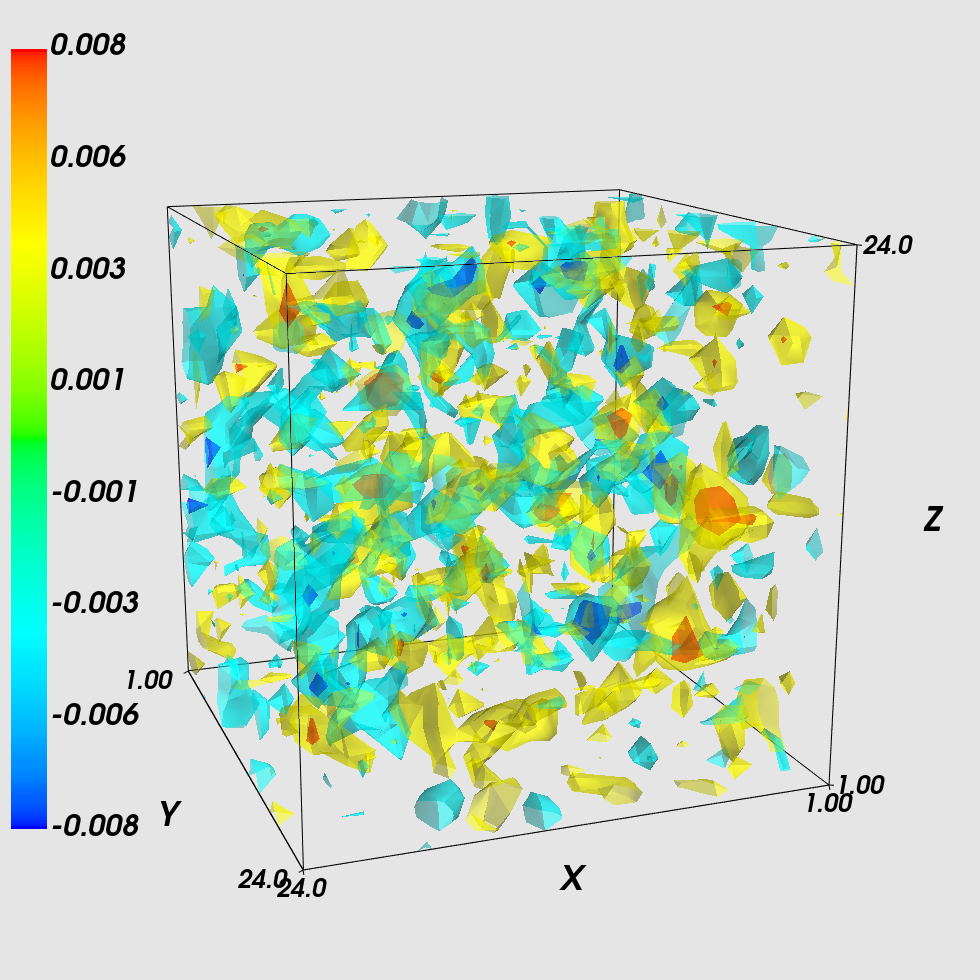}\\
			\centerline{$\kappa=0.19$}
		\end{minipage}
		\begin{minipage}[t]{0.36\textwidth}
			\includegraphics[width=2.25in]{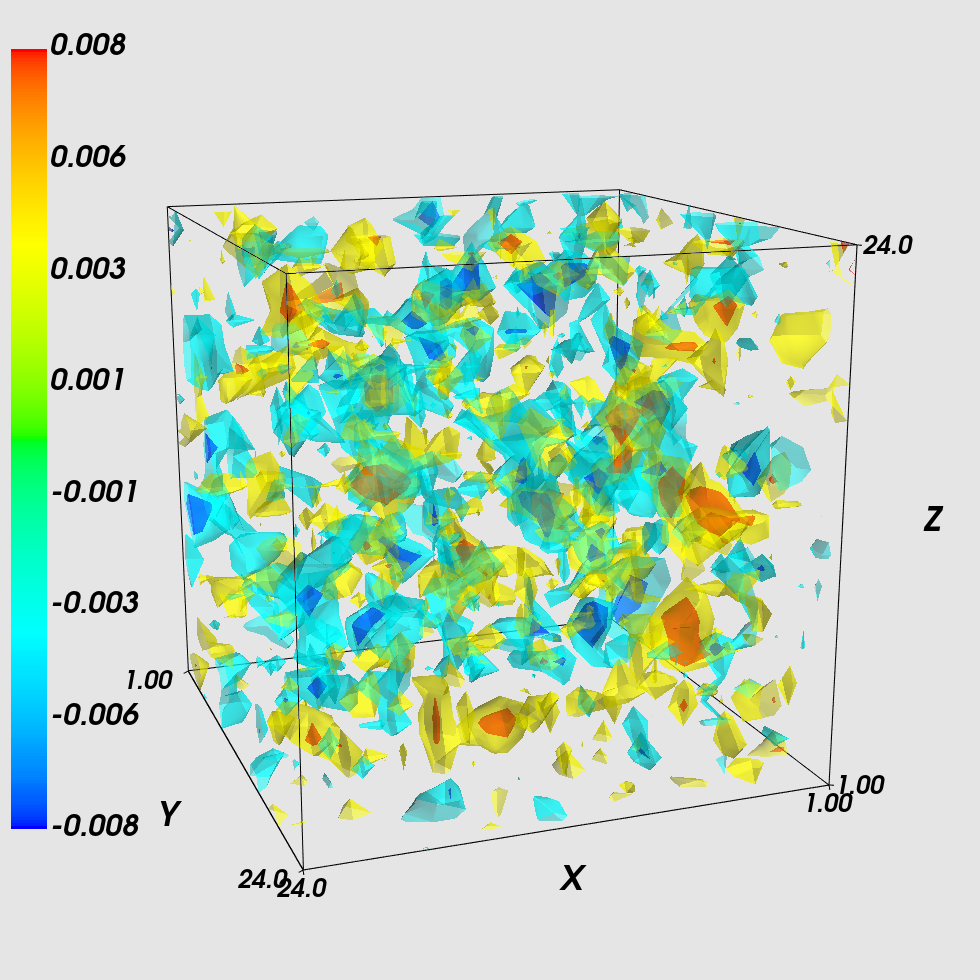}\\
			\centerline{$\tau_{{\rm pr}}=0.325$}
	\end{minipage}	 }
	\hspace{-15mm}
	\subfigure{
		\begin{minipage}[t]{0.36\textwidth}
			\includegraphics[width=2.25in]{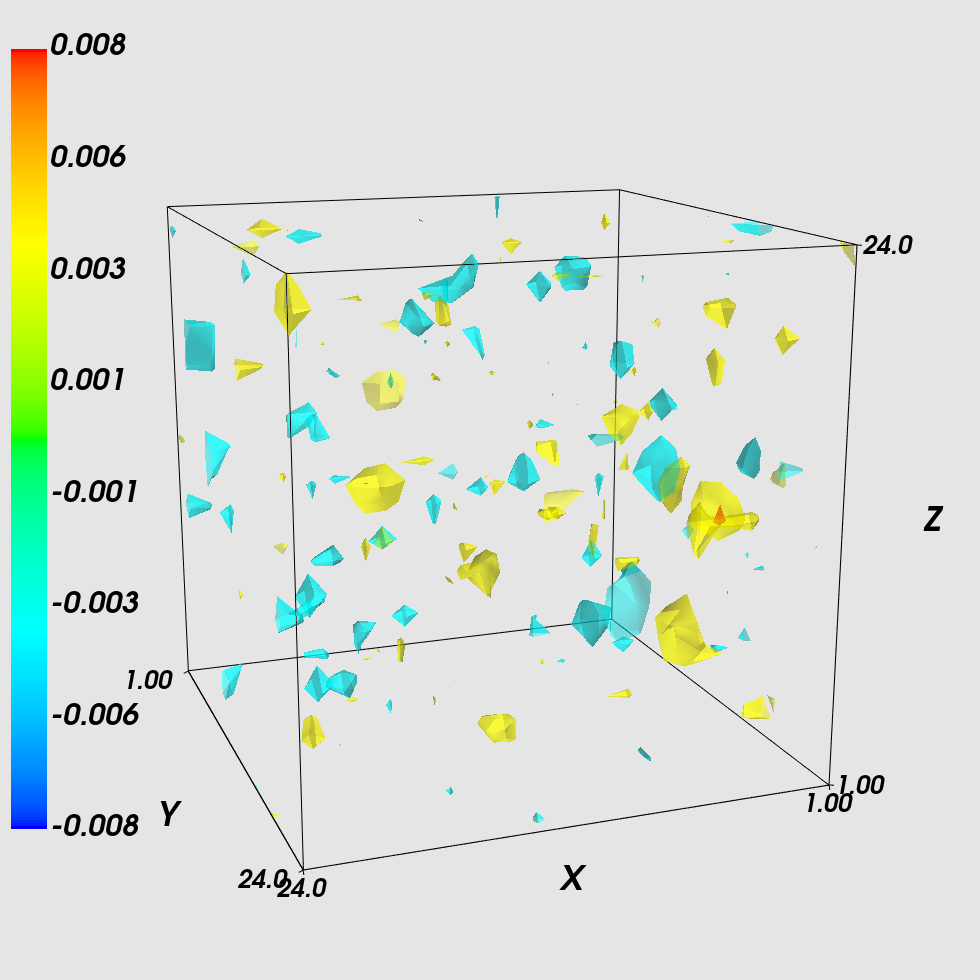}\\
			\centerline{$\kappa=0.17$}
		\end{minipage}
		\begin{minipage}[t]{0.36\textwidth}
			\includegraphics[width=2.25in]{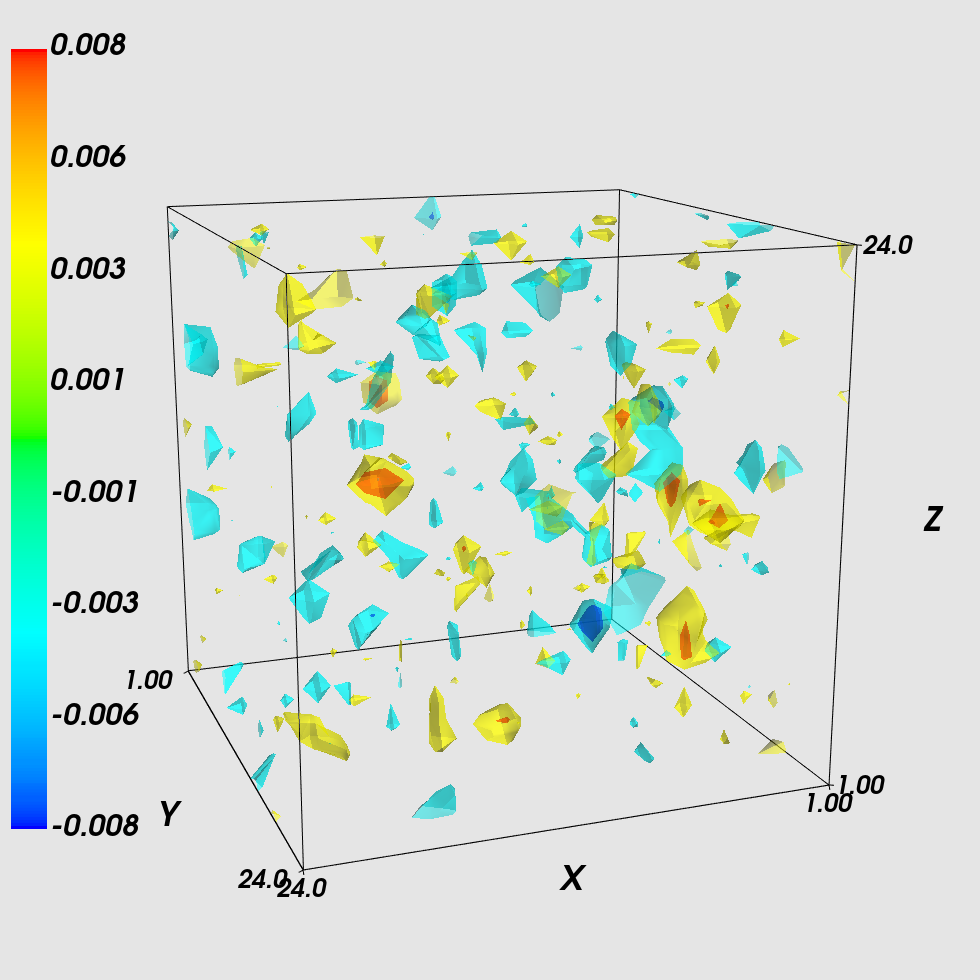}\\
			\centerline{$\tau_{{\rm pr}}=0.365$}
	\end{minipage}	 }
	\caption{The best matched topological charge density $q_{\text{wf}}\left(x\right)$ (right) calculated by the Wilson flow method compared with the overlap $q_{\text{smp}}\left(x\right)$  for the time slice $t = 24$, where $q_{\text{wf}}\left(x\right)$ is renormalized using $Z_{\rm {calc}}$. $n_{\rm{pr}}$ is the proper number of Wilson flow fixed by matching procedure. A color cutoff method is used shown as the color map in the figures. \label{fig:best-matched}}
\end{figure}

The SMP topological charge density for three choices $\kappa$ compared
with the proper number of Wilson flow $\tau_{{\rm pr}}$ for time slice $t=24$ as examples are shown in Fig. \ref{fig:best-matched}, and $q_{\text{wf}}\left(x\right)$ is renormalized using $Z_{\rm {calc}}$. Other time slices have the similar property. It shows that more Wilson flow time for $q_{\text{wf}}\left(x\right)$
are needed to match the topological charge density $q_{\text{smp}}\left(x\right)$
with a smaller $\kappa$ or larger mass $m$. This phenomenon may be owing to that the samller $\kappa$ shows more sparse small eigenvalues of $D_{\rm ov}$, which has the similar effect of smoothing the configurations. However, the detailed reasons need further study. It indicates that the overlap Dirac operator
is less sensitive to small objects as $\kappa$ is decreased, and these objects can be removed by Wilson flow smoothing.

$\tau_{\rm pr}$, $Z_{\text{calc}}$, $\Xi_{AB}$, $Q_{\rm smp}$ and $Q_{\rm wf}$ of different ensembles for five different $\kappa$ are shown in Table \ref{tab:table-x16t16-conf1}$ \sim$ \ref{tab:table-x32t32-conf1}, respectively. When calculating $Z_{{\rm calc}}$, the flow time of Wilson flow is $\tau_{\rm pr}$. As the parameter $\kappa$ increases, there is a monotonically decreasing trend in the proper flow time of Wilson flow in all tables.  It shows that the matching procedure is effective and the proper flow time of Wilson flow is almost equal at the same $\kappa$ for different configurations of the same lattice ensembles. When the parameter $\kappa$ is fixed, $\Xi_{\rm{AB}}$ and $Z_{\text{calc}}$ are independent of topological charge $Q$ and very close for different configurations with the same lattice volume. The value of $\Xi_{\rm{AB}}$ is approximately in the range $\left[0.68,0.77\right]$. We can see that the proper flow time of Wilson flow $\tau_{\rm pr}$ is different for different $\kappa$. However, it is reasonable to choose the average value of the proper flow time of Wilson flow of different $\kappa$ as the proper $\bar{\tau}_{\rm pr}$. And it is about $\bar{\tau}_{\rm pr}=0.345$, $\bar{\tau}_{\rm pr}=0.327$, $\bar{\tau}_{\rm pr}=0.317$ or the proper flow radius of Wilson flow  $\sqrt{8\tau} \approx 0.214$, $0.137$ and $0.104\thinspace{\text{fm}}$
 for lattice ensemble $16^{4}$ at $\beta=4.5$, $24^{3}\times48$ at $\beta=4.8$ and $32^{4}$ at $\beta=5.0$, respectively. It indicates that we can choose $\bar{\tau}_{{\rm pr}}$ for gluonic $q_{{\rm wf}}\left(x\right)$ to match with fermionic $q_{{\rm smp}}\left(x\right)$.

All results show that $Q_{\rm wf}$ deviates largely from an integer for the proper flow time of Wilson flow fixed by the matching procedure. This phenomenon may owe to that the topological charge is the global property of topological structure. However, the matching parameter $\Xi_{AB}$ only shows the local matching property of topological charge density. Otherwise, the topological charge is the effect of the infrared properties, and the ultraviolet fluctuations lead to unphysical results as well as to non-integer topological charge values. Wilson flow is indeed a smoothing scheme of gauge fields. But Wilson flow will modify the gauge field at the same time, which is not justified that the topological charge is conserved. On the other hand, $F_{\mu\nu}^{\text{Imp}}$ in Eq.(\ref{q-impro}) is applicable under the classical expansion with respect to lattice spacing $a$, and it may affected by some quantum fluctuations even though the smoothing is performed. 
		
However, the topological charge is not the main quantity of interest, and the physically relevant observable is the topological susceptibility. The results show that in order to make the topological charge close to an integer, larger Wilson flow time is required. It is note that too large Wilson flow time may wipe out the negative core of topological charge density correlator\cite{Chowdhury:2014mra}. However, the flow time for the topological susceptibility reaches a plateau is smaller than the flow time for the topological charges reach some integers \cite{Alexandrou:2015yba}. But the topological susceptibility by using SMP method is too expensive to calculated. In the future work, we may try to consider it.

\begin{table}[H]
	\centering
	\begin{tabular}{p{2.3cm}p{2.3cm}p{2.3cm}p{2.3cm}p{2.3cm}p{2.3cm}p{2.3cm}}
		\hline
		\makecell[c]{$\kappa$} &   \makecell[c]{$\tau_{{\rm pr}}$} & \makecell[c]{$Z_{{\rm calc}}$} & \makecell[c]{$\Xi_{{\rm AB}}$} & \makecell[c]{$Q_{{\rm smp}}$} & \makecell[c]{$Q_{\rm wf}$}\tabularnewline
		\hline
		\makecell[c]{$0.17$} &  \makecell[c]{$0.360$} & \makecell[c]{$0.5121$} & \makecell[c]{$0.6853$} & \makecell[c]{$5.0010$} & \makecell[c]{3.5420}\tabularnewline
		\makecell[c]{$0.18$} &  \makecell[c]{$0.365$} & \makecell[c]{$0.7351$} & \makecell[c]{$0.7259$} & \makecell[c]{$5.0015$} & \makecell[c]{3.5299}\tabularnewline
		\makecell[c]{$0.19$} &  \makecell[c]{$0.350$} & \makecell[c]{$0.8656$} & \makecell[c]{$0.7309$} & \makecell[c]{$6.0005$} & \makecell[c]{3.5847}\tabularnewline
		\makecell[c]{$0.21$} &  \makecell[c]{$0.330$} & \makecell[c]{$1.0045$} & \makecell[c]{$0.7267$} & \makecell[c]{$3.9990$} & \makecell[c]{3.6378}\tabularnewline
		\makecell[c]{$0.23$} &  \makecell[c]{$0.320$} & \makecell[c]{$1.0073$} & \makecell[c]{$0.7030$} & \makecell[c]{$1.9963$} & \makecell[c]{3.6796}\tabularnewline
		\hline
	\end{tabular}
	\caption{The proper time of Wilson flow, $\tau_{\rm pr}$, needed to match the SMP
		topological charge density with different $\kappa$ at $\beta=4.50$ and lattice volume $16^{4}$ for  ${\rm conf.}\thinspace1$. $Q_{\rm smp}$ is obtained by SMP method, and $Q_{\rm wf}$ is the result of Wilson flow with $\tau_{\rm pr}$. \label{tab:table-x16t16-conf1} }
\end{table}

\begin{table}[H]
	\centering
	\begin{tabular}{p{2.3cm}p{2.3cm}p{2.3cm}p{2.3cm}p{2.3cm}p{2.3cm}p{2.3cm}}
		\hline
		\makecell[c]{$\kappa$} &   \makecell[c]{$\tau_{{\rm pr}}$} & \makecell[c]{$Z_{{\rm calc}}$} & \makecell[c]{$\Xi_{{\rm AB}}$} & \makecell[c]{$Q_{{\rm smp}}$} & \makecell[c]{$Q_{\rm wf}$}\tabularnewline
		\hline
		\makecell[c]{$0.17$} &  \makecell[c]{$0.365$} & \makecell[c]{$0.5230$} & \makecell[c]{$0.7006$} & \makecell[c]{$-6.0091$} & \makecell[c]{-7.2945}\tabularnewline
		\makecell[c]{$0.18$} &  \makecell[c]{$0.360$} & \makecell[c]{$0.7235$} & \makecell[c]{$0.7339$} & \makecell[c]{$-6.0052$} & \makecell[c]{-7.3027}\tabularnewline
		\makecell[c]{$0.19$} &  \makecell[c]{$0.350$} & \makecell[c]{$0.8680$} & \makecell[c]{$0.7471$} & \makecell[c]{$-5.0026$} & \makecell[c]{-7.3181}\tabularnewline
		\makecell[c]{$0.21$} &  \makecell[c]{$0.330$} & \makecell[c]{$1.0077$} & \makecell[c]{$0.7322$} & \makecell[c]{$-3.9997$} & \makecell[c]{-7.3458}\tabularnewline
		\makecell[c]{$0.23$} &  \makecell[c]{$0.320$} & \makecell[c]{$1.0121$} & \makecell[c]{$0.7096$} & \makecell[c]{$-5.9964$} & \makecell[c]{-7.3581}\tabularnewline
		\hline
	\end{tabular}
	\caption{The proper time of Wilson flow, $\tau_{\rm pr}$, needed to match the SMP
		topological charge density with different $\kappa$ at $\beta=4.50$ and lattice volume $16^{4}$ for ${\rm conf.}\thinspace2$. $Q_{\rm smp}$ is obtained by SMP method, and $Q_{\rm wf}$ is the result of Wilson flow with $\tau_{\rm pr}$. \label{tab:table-x16t16-conf2}}
\end{table}

\begin{table}[H]
	\centering
	\begin{tabular}{p{2.3cm}p{2.3cm}p{2.3cm}p{2.3cm}p{2.3cm}p{2.3cm}p{2.3cm}}
		\hline
		\makecell[c]{$\kappa$} &   \makecell[c]{$\tau_{{\rm pr}}$} & \makecell[c]{$Z_{{\rm calc}}$} & \makecell[c]{$\Xi_{{\rm AB}}$} & \makecell[c]{$Q_{{\rm smp}}$} & \makecell[c]{$Q_{\rm wf}$} \tabularnewline
		\hline
		\makecell[c]{$0.17$} &  \makecell[c]{$0.365$} & \makecell[c]{$0.6811$} & \makecell[c]{$0.7653$} & \makecell[c]{$8.0077$} & \makecell[c]{8.5997}\tabularnewline
		\makecell[c]{$0.18$} &  \makecell[c]{$0.345$} & \makecell[c]{$0.8364$} & \makecell[c]{$0.7670$} & \makecell[c]{$7.0078$} & \makecell[c]{8.6863}\tabularnewline
		\makecell[c]{$0.19$} &  \makecell[c]{$0.325$} & \makecell[c]{$0.9275$} & \makecell[c]{$0.7580$} & \makecell[c]{$7.0095$} & \makecell[c]{8.7954}\tabularnewline
		\makecell[c]{$0.21$} &  \makecell[c]{$0.305$} & \makecell[c]{$1.0215$} & \makecell[c]{$0.7324$} & \makecell[c]{$9.0169$} & \makecell[c]{8.9322}\tabularnewline
		\makecell[c]{$0.23$} &  \makecell[c]{$0.295$} & \makecell[c]{$0.9811$} & \makecell[c]{$0.6972$} & \makecell[c]{$9.0279$} & \makecell[c]{9.0130}\tabularnewline
		\hline
	\end{tabular}
	
	\caption{The proper time of Wilson flow, $\tau_{\rm pr}$, needed to match the SMP
		topological charge density with different $\kappa$ at $\beta=4.80$ and lattice volume $24^{3}\times48$ for ${\rm conf.}\thinspace1$. $Q_{\rm smp}$ is obtained by SMP method, and $Q_{\rm wf}$ is the result of Wilson flow with $\tau_{\rm pr}$. \label{tab:table-x24t48-conf1}}
\end{table}

\begin{table}[H]
	\centering
	\begin{tabular}{p{2.3cm}p{2.3cm}p{2.3cm}p{2.3cm}p{2.3cm}p{2.3cm}p{2.3cm}}
		\hline
		\makecell[c]{$\kappa$} &   \makecell[c]{$\tau_{{\rm pr}}$} & \makecell[c]{$Z_{{\rm calc}}$} & \makecell[c]{$\Xi_{{\rm AB}}$} & \makecell[c]{$Q_{{\rm smp}}$} & \makecell[c]{$Q_{\rm wf}$} \tabularnewline
		\hline
		\makecell[c]{$0.17$} &  \makecell[c]{$0.365$} & \makecell[c]{$0.6802$} & \makecell[c]{$0.7661$} & \makecell[c]{$4.9938$} & \makecell[c]{4.3224}\tabularnewline
		\makecell[c]{$0.18$} &  \makecell[c]{$0.345$} & \makecell[c]{$0.8359$} & \makecell[c]{$0.7656$} & \makecell[c]{$4.9932$} & \makecell[c]{4.3289}\tabularnewline
		\makecell[c]{$0.19$} &  \makecell[c]{$0.325$} & \makecell[c]{$0.9274$} & \makecell[c]{$0.7574$} & \makecell[c]{$3.9942$} & \makecell[c]{4.3365}\tabularnewline
		\makecell[c]{$0.21$} &  \makecell[c]{$0.305$} & \makecell[c]{$1.0211$} & \makecell[c]{$0.7309$} & \makecell[c]{$3.9949$} & \makecell[c]{4.3431}\tabularnewline
		\makecell[c]{$0.23$} &  \makecell[c]{$0.290$} & \makecell[c]{$0.9596$} & \makecell[c]{$0.6953$} & \makecell[c]{$3.9967$} & \makecell[c]{4.3454}\tabularnewline
		\hline
	\end{tabular}
	
	\caption{The proper time of Wilson flow, $\tau_{\rm pr}$, needed to match the SMP
		topological charge density with varied $\kappa$ at $\beta=4.80$ and lattice volume $24^{3}\times48$ for ${\rm conf.}\thinspace2$. $Q_{\rm smp}$ is obtained by SMP method, and $Q_{\rm wf}$ is the result of Wilson flow with $\tau_{\rm pr}$. \label{tab:table-x24t48-conf2}}
\end{table}

\begin{table}[H]
	\centering
	\begin{tabular}{p{2.3cm}p{2.3cm}p{2.3cm}p{2.3cm}p{2.3cm}p{2.3cm}p{2.3cm}}
		\hline
		\makecell[c]{$\kappa$} &   \makecell[c]{$\tau_{{\rm pr}}$} & \makecell[c]{$Z_{{\rm calc}}$} & \makecell[c]{$\Xi_{{\rm AB}}$} & \makecell[c]{$Q_{{\rm smp}}$} & \makecell[c]{$Q_{\rm wf}$}\tabularnewline
		\hline
		\makecell[c]{$0.17$} &  \makecell[c]{$0.355$} & \makecell[c]{$0.7271$} & \makecell[c]{$0.7678$} & \makecell[c]{$3.0099$} & \makecell[c]{2.8066}\tabularnewline
		\makecell[c]{$0.18$} &  \makecell[c]{$0.335$} & \makecell[c]{$0.8724$} & \makecell[c]{$0.7627$} & \makecell[c]{$3.0059$} & \makecell[c]{2.7340}\tabularnewline
		\makecell[c]{$0.19$} &  \makecell[c]{$0.315$} & \makecell[c]{$0.9507$} & \makecell[c]{$0.7515$} & \makecell[c]{$3.0069$} & \makecell[c]{2.6380}\tabularnewline
		\makecell[c]{$0.21$} &  \makecell[c]{$0.295$} & \makecell[c]{$1.0233$} & \makecell[c]{$0.7222$} & \makecell[c]{$3.0127$} & \makecell[c]{2.5120}\tabularnewline
		\makecell[c]{$0.23$} &  \makecell[c]{$0.285$} & \makecell[c]{$0.9636$} & \makecell[c]{$0.6818$} & \makecell[c]{$3.0240$} & \makecell[c]{2.4352}\tabularnewline
		\hline
	\end{tabular}
	
	\caption{The proper time of Wilson flow, $\tau_{\rm pr}$, needed to match the SMP
		topological charge density with varied $\kappa$ at $\beta=5.0$ and lattice volume $32^{4}$. $Q_{\rm smp}$ is obtained by SMP method, and $Q_{\rm wf}$ is the result of Wilson flow with $\tau_{\rm pr}$. \label{tab:table-x32t32-conf1}}
\end{table}

\begin{figure}[H]
	\centering	
	\includegraphics[scale=0.22]{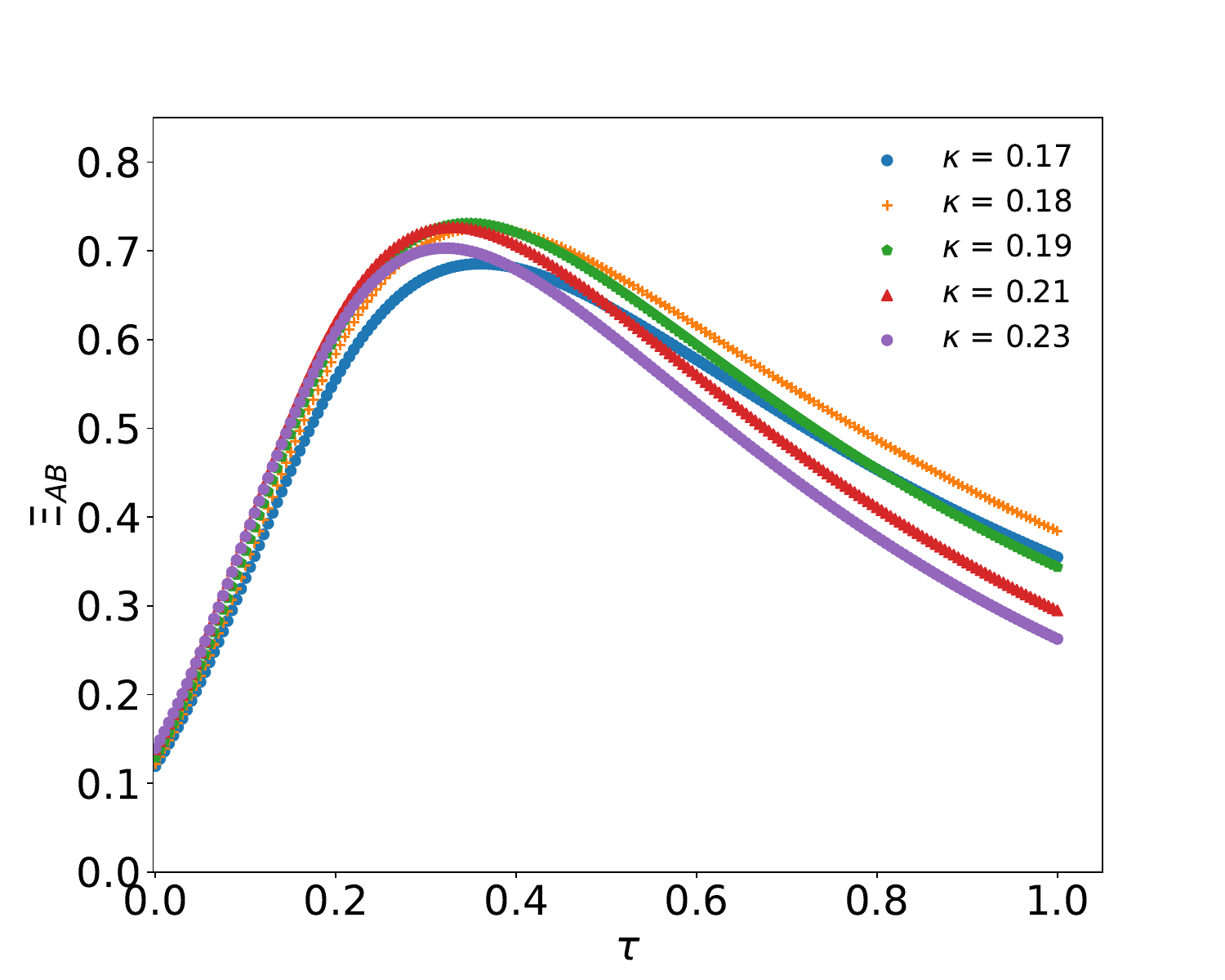}\includegraphics[scale=0.22]{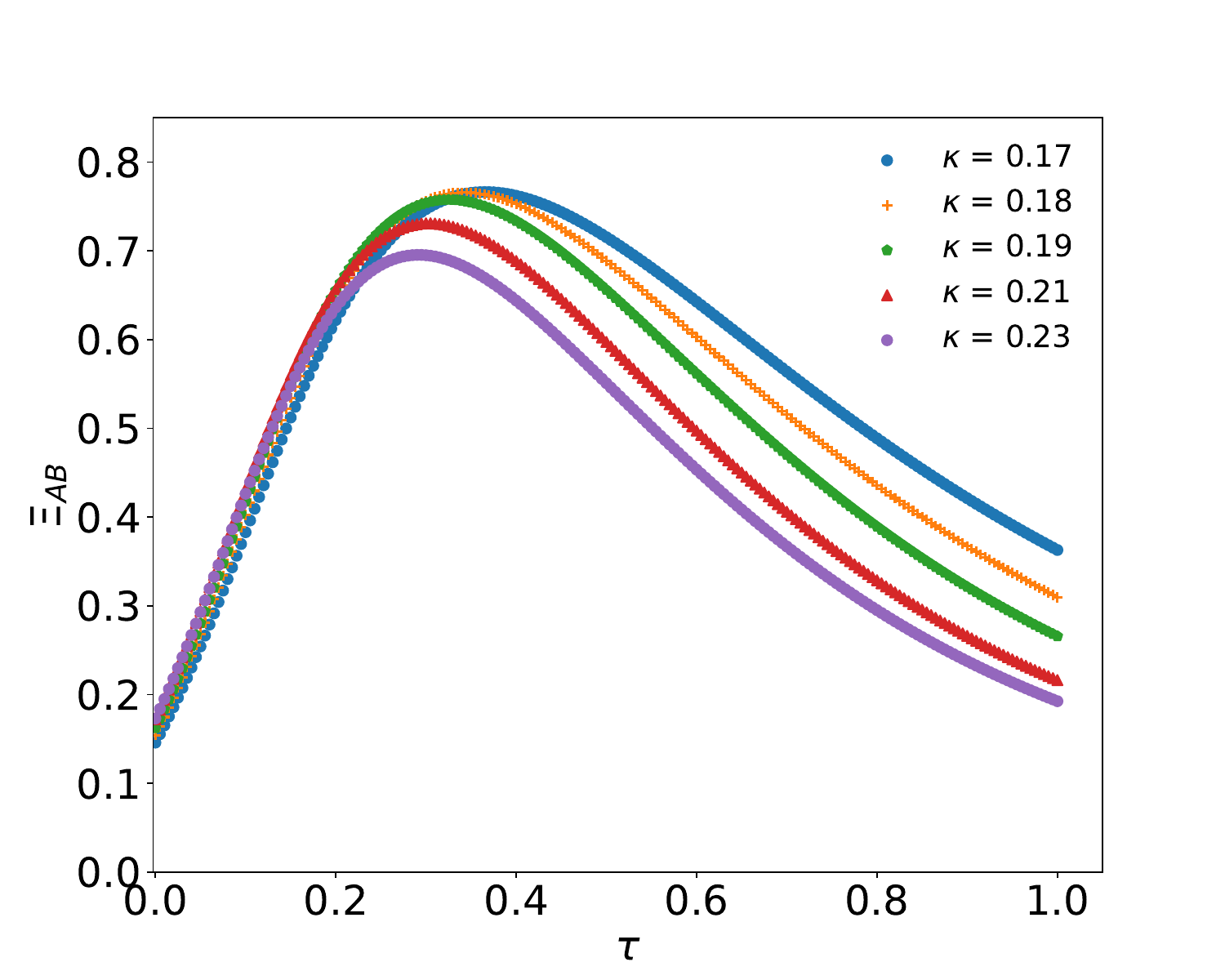}\includegraphics[scale=0.22]{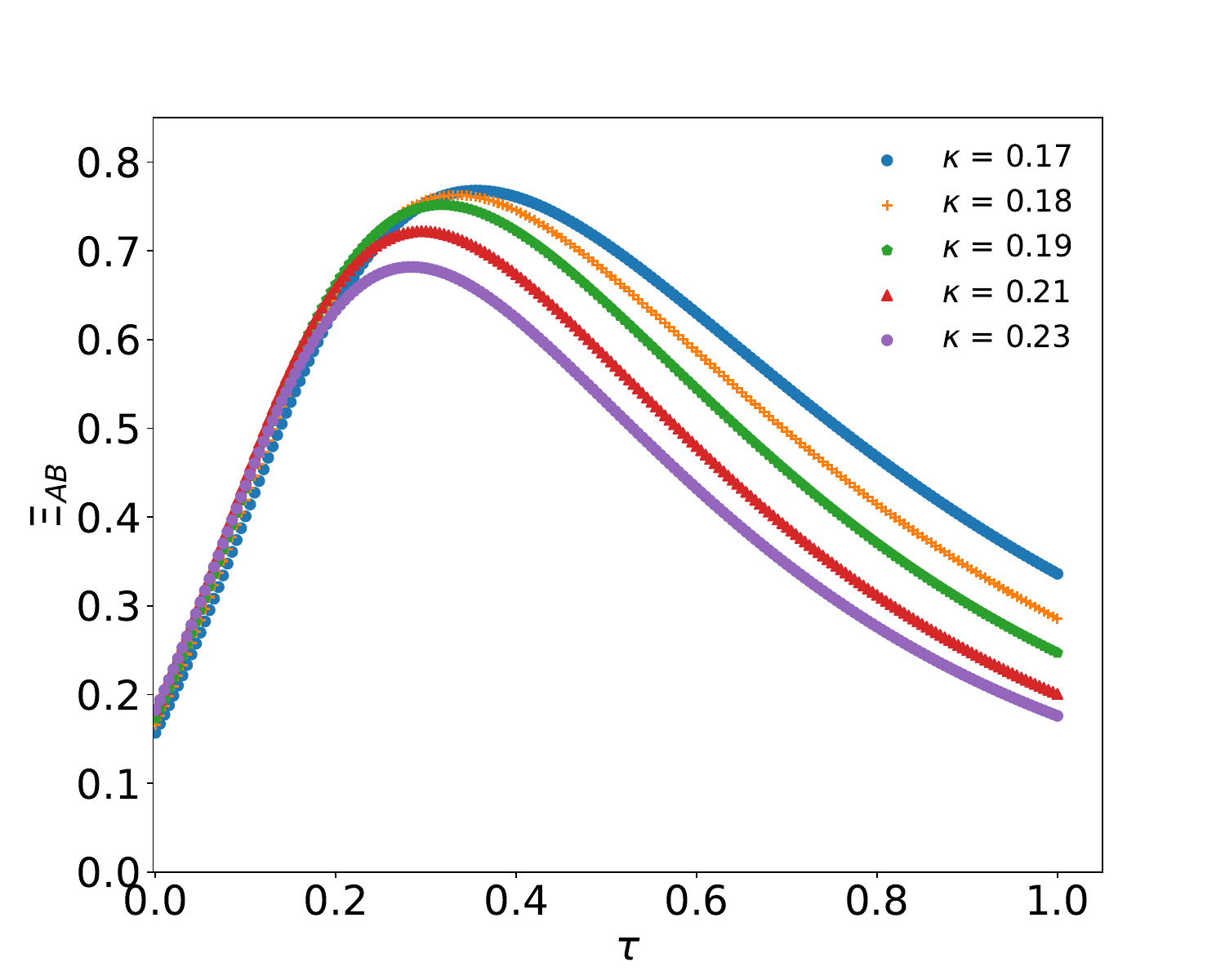}
	\caption{$\Xi_{{\rm AB}}$ versus $\tau$ of one configuration for lattices of $16^4$, $24^{3}\times48$ and $32^4$. From left to right, the first figure is the $\Xi_{\rm {AB}}$ versus $\tau$ for lattice volume $16^{4}$, the second figure for lattice volume $24^{3}\times48$, and the last one for lattice volume $32^{4}$. As $\kappa$ is reduced, the proper flow time of Wilson flow is increased. As the lattice spacing decreases, $\Xi_{{\rm AB}}$ tends to increase. \label{fig:XI_AB}}	
\end{figure}

In Fig. \ref{fig:XI_AB},  $\Xi_{\text{AB}}$ versus the flow time of Wilson flow of one configuration for lattices of $16^4$, $24^{3}\times48$ and $32^4$ are shown. $\Xi_{\text{AB}}$ for other configurations have the similar trend. We see that as the flow time of Wilson flow $\tau$
increases, $\Xi_{{\rm AB}}$ reaches to a maximum value and then decreases. When the parameter $\kappa$ is fixed, we can observe that as the lattice spacing decreases, the proper flow time of Wilson flow almost uniformly decreases, as expected. It also shows that as the lattice spacing $a$ decreases, $\Xi_{\text{AB}}$ tends to increase.

\section{Conclusions}

We have analyzed the topological charge density $q\left(x\right)$ of all time slices
using direct visualizations. We find that the SMP method is a good choice
to study the topological charge density in the fermionic definition. And the SMP method is much cheaper than the traditional point source method, especially for large lattice volume. The results show that the topological
charge density depends on the Wilson mass parameter $m$ in the fermionic definition. By comparing the $q_{{\rm smp}}\left(x\right)$ with the gluonic definition of $q_{{\rm wf}}\left(x\right)$,
a correlation between $m$ and $\tau$ is revealed. Smaller
values of $\kappa$ remove non-trivial topological charge fluctuations, which are similar to Wilson flow with a larger flow time. The detailed reasons are worthy of further study. By analyzing the topological
charge density $q\left(x\right)$, we find that the proper flow time of Wilson flow
for the gluonic definition of topological charge density can be obtained by the comparison of  $q_{{\rm smp}}\left(x\right)$ with $q_{{\rm wf}}\left(x\right)$. We also observe that the proper flow time of Wilson flow decreases with the decrease of lattice spacing $a$, which is consistent with expectations. Furthermore, as the lattice spacing $a$ decreases, $\Xi_{\text{AB}}$ tends to increase. 

We also find that topological charge obtained by Wilsonf flow at the proper flow time is far away from integer, and it is different from that of fermonic definition. Although the topological charge density of fermionic definition and that of gluonic definition has the best match, the topological charge from fermonic and gluonic definition are very different. The reason of phenomenon may be that the matching parameter $\Xi_{AB}$ only shows the local matching property of topological charge density. But the topological charge is the global property of topological structure. On the other hand, Wilson flow indeed can smooth the gauge field. But Wilson flow also changes the configurations, which does not guarantee the conservation of topological charge. Otherwise, in the gluonic definition of topological charge density, we used the improved field strength tensor corrected in the classical expansion with respect to lattice spacing and may be affected by the quantum fluctuations even though the Wilson flow is used. The detailed reasons need further study.  

In the SMP method, we had known that the error is dependent on the off-diagnal components of overlap operator. With larger distance parameter, it has smaller errors in SMP method. We can try to choose larger distance parameter to decrease the error in the future work. Otherwise, we could try to improve the field strenth tensor to reduce the effect of classical expansion with respect to the lattice spacing in gluonic definition of topological charge density. 

\acknowledgments
We are grateful to Heng-tong Ding for careful reading of the manuscript and useful suggestions.
Numerical simulations have been performed on the Tianhe-2 supercomputer at the National Supercomputer Center in Guangzhou(NSCC-GZ), China. This work is supported by the National Natural Science Foundation of China (NSFC) under the project No. 11335001.

\normalem

\providecommand{\href}[2]{#2}\begingroup\raggedright\endgroup


\begin{thebibliography}{99}
	
	\bibitem{Schierholz:1994pb}
	G.~Schierholz, {\it
		{Towards a dynamical solution of the strong CP problem}},  {\em Nucl. Phys. Proc. Suppl.} {\bf 37A}
	(1994) 203--210 [\href{http://arXiv.org/abs/hep-lat/9403012}{{\tt
			hep-lat/9403012}}].
	
	\bibitem{Witten:1978bc}
	Edward Witten, {\it
		{Instantons, the Quark Model, and the 1/N Expansion}},  {\em Nucl. Phys. B} {\bf 149}
	(1979) 285--320.
	
	\bibitem{Diakonov:1995ea}
	Dmitri Diakonov, {\it
		{Chiral symmetry breaking by instantons}},  {\em Proc. Int. Sch. Phys. Fermi} {\bf 130}
	(1996) 397--432 [\href{http://arXiv.org/abs/hep-ph/9602375}{{\tt
			hep-ph/9602375}}].
	
	\bibitem{Cichy:2014qta}
	Krzysztof Cichy, Arthur Dromard, Elena Garcia-Ramos, Konstantin Ottnad, Carsten
	Urbach, Marc Wagner, Urs Wenger, and Falk Zimmermann, {\it
		{Comparison of different lattice definitions of the topological
			charge}},  {\em PoS} {\bf LATTICE2014}
	(2014) 075 [\href{http://arXiv.org/abs/1411.1205}{{\tt
			1411.1205}}].
	
	\bibitem{Muller-Preussker:2015daa}
	M.~M{\"u}ller-Preussker, {\it
		{Recent results on topology on the lattice (in memory of Pierre van
			Baal)}},  {\em PoS} {\bf LATTICE2014}
	(2015) 003 [\href{http://arXiv.org/abs/1503.01254}{{\tt
			1503.01254}}].	
	
	\bibitem{Alexandrou:2017hqw}
	Constantia Alexandrou, Andreas Athenodorou, Krzysztof Cichy, Arthur Dromard,
	Elena Garcia-Ramos, Karl Jansen, Urs Wenger, and Falk Zimmermann, {\it
		{Comparison of topological charge definitions in Lattice QCD}},  {\em Eur. Phys. J. C} {\bf 80}
	(2020) 424 [\href{http://arXiv.org/abs/1708.00696}{{\tt
			1708.00696}}].	
	
	\bibitem{Atiyah:1971rm}
	M.~F. Atiyah and I.~M. Singer, {\it
		{The Index of elliptic operators. 5.}},  {\em Annals Math.} {\bf 93}
	(1971) 139--149.
	
	\bibitem{Hasenfratz:1998ri}
	Peter Hasenfratz, Victor Laliena, and Ferenc Niedermayer, {\it
		{The Index theorem in QCD with a finite cutoff}},  {\em Phys. Lett. B} {\bf 427}
	(1998) 125--131 [\href{http://arXiv.org/abs/hep-lat/9801021}{{\tt
			hep-lat/9801021}}].
	
	\bibitem{Belavin:1975fg}
	A.A. Belavin, Alexander~M. Polyakov, A.S. Schwartz, and Yu.S. Tyupkin, {\it
		{Pseudoparticle Solutions of the Yang-Mills Equations}},  {\em Phys. Lett. B} {\bf 59}
	(1975) 85--87.
	
	\bibitem{Fujikawa:1998if}
	Kazuo Fujikawa, {\it
		{A continuum limit of the chiral jacobian in lattice gauge theory}},  {\em Nucl. Phys. B} {\bf 546}
	(1999) 480--494.
	
	
	\bibitem{Kikukawa:1998pd}
	Yoshio Kikukawa and Atsushi Yamada, {\it
		{Weak coupling expansion of massless QCD with a Ginsparg-Wilson
			fermion and axial U(1) anomaly}},  {\em Phys. Lett. B} {\bf 448}
	(1999) 265--274 [\href{http://arXiv.org/abs/hep-lat/9806013}{{\tt
			hep-lat/9806013}}].
	
	\bibitem{Neuberger:1997fp}
	Herbert Neuberger, {\it
		{Exactly massless quarks on the lattice}},  {\em Phys. Lett. B} {\bf 417}
	(1998) 141--144 [\href{http://arXiv.org/abs/hep-lat/9707022}{{\tt
			hep-lat/9707022}}].
	
	\bibitem{Neuberger:1998wv}
	Herbert Neuberger, {\it
		{More about exactly massless quarks on the lattice}},  {\em Phys. Lett. B} {\bf 427}
	(1998) 353--355 [\href{http://arXiv.org/abs/hep-lat/9801031}{{\tt
			hep-lat/9801031}}].
	
	\bibitem{Horvath:2003yj}
	I.Horv\'ath, S.J. Dong, Terrence Draper, F.X. Lee, K.F. Liu, N.~Mathur, H.B.
	Thacker, and J.B. Zhang, {\it
		{Low dimensional long range topological charge structure in the QCD
			vacuum}},  {\em Phys. Rev. D} {\bf 68}
	(2003) 114505 [\href{http://arXiv.org/abs/hep-lat/0302009}{{\tt
			hep-lat/0302009}}].
	
	\bibitem{Ilgenfritz:2007xu}
	E.-M. Ilgenfritz, K.~Koller, Y.~Koma, G.~Schierholz, T.~Streuer, and
	V.~Weinberg, {\it
		{Exploring the structure of the quenched QCD vacuum with overlap
			fermions}},  {\em Phys. Rev. D} {\bf 76}
	(2007) 034506 [\href{http://arXiv.org/abs/0705.0018}{{\tt 0705.0018}}].
	
	\bibitem{Xiong:2019pmh}
	Guang-Yi Xiong, Jian-Bo Zhang, and You-Hao Zou, {\it
		{Evaluating the topological charge density with the symmetric
			multi-probing method}},  {\em Chin. Phys. C} {\bf 43(3)}
	(2019) 033102.
	
	\bibitem{Narayanan1995}
	Rajamani Narayanan and Herbert Neuberger, {\it
		{A Construction of lattice chiral gauge theories}},  {\em Nucl. Phys. B} {\bf 443}
	(1995) 305--385 [\href{http://arXiv.org/abs/hep-th/9411108}{{\tt
			hep-th/9411108}}].
	
	\bibitem{Edwards:1998sh}
	Robert~G. Edwards, Urs~M. Heller, and Rajamani Narayanan, {\it
		{Spectral flow, chiral condensate and topology in lattice QCD}},  {\em Nucl. Phys. B} {\bf 535}
	(1998) 403--422 [\href{http://arXiv.org/abs/hep-lat/9802016}{{\tt
			hep-lat/9802016}}].
	
	\bibitem{Narayanan:1997sa}
	Rajamani Narayanan and Pavlos~M. Vranas, {\it
		{A Numerical test of the continuum index theorem on the lattice}},  {\em Nucl. Phys. B} {\bf 506}
	(1997) 373--386 [\href{http://arXiv.org/abs/hep-lat/9702005}{{\tt
			hep-lat/9702005}}].
	
	\bibitem{Zhang:2001fk}
	J.B. Zhang, S.O. Bilson-Thompson, F.D.R. Bonnet, D.B. Leinweber, Anthony~G.
	Williams, and J.M. Zanotti, {\it
		{Numerical study of lattice index theorem using improved cooling and
			overlap fermions}},  {\em Phys. Rev. D} {\bf 65}
	(2002) 074510 [\href{http://arXiv.org/abs/hep-lat/0111060}{{\tt
			hep-lat/0111060}}].
	
	\bibitem{Vege:2019nee}
	Hans Mathias~Mamen Vege, {\it
		{Solving SU(3) Yang-Mills theory on the lattice: a calculation of
			selected gauge observables with gradient flow}},  {\em Master's thesis} (2019) U. Oslo (main).
	
	\bibitem{Moran:2010rn}
	Peter~J. Moran, Derek~B. Leinweber, and Jianbo Zhang, {\it
		{Wilson mass dependence of the overlap topological charge density}},  {\em Phys. Lett. B} {\bf 695}
	(2011) 337--342 [\href{http://arXiv.org/abs/1007.0854}{{\tt
			1007.0854}}].
	
	\bibitem{Luscher:1984xn}
	M.~L\"{u}scher and P.~Weisz, {\it
		{On-Shell Improved Lattice Gauge Theories}},  {\em Commun. Math. Phys.} {\bf 97}
	(1985) 59 [Erratum: Commun. Math. Phys. 98 (1985) 433].
	
	\bibitem{Bonnet:2001rc}
	Frederic~D.R. Bonnet, Derek~B. Leinweber, Anthony~G. Williams, and James~M.
	Zanotti, {\it
		{Improved smoothing algorithms for lattice gauge theory}},  {\em Phys. Rev. D} {\bf 65}
	(2002) 114510 [\href{http://arXiv.org/abs/hep-lat/0106023}{{\tt
			hep-lat/0106023}}].
	
	\bibitem{Cheng:2020dlb}
	Zhen Cheng, Jian-Bo Zhang, and Guang-Yi Xiong, {\it
		{Calculation of disconnected quark loops in lattice QCD}},  {\em Chin. Phys. C} {\bf 44(3)}
	(2020) 033104.
	
	
	\bibitem{Luscher:2010iy}
	Martin L\"{u}scher, {\it
		{Properties and uses of the Wilson flow in lattice QCD}},  {\em JHEP} {\bf 08}
	(2010) 071 [Erratum: {\em JHEP} {\bf 03} (2014) 092]. [\href{http://arXiv.org/abs/1006.4518}{{\tt
			1006.4518}}].
		
	 \bibitem{Bonati:2014tqa}
	Claudio Bonati and Massimo D'Elia, {\it
		{Comparison of the gradient flow with cooling in $SU(3)$ pure gauge theory}},  {\em Phys. Rev. D} {\bf 89}
	(2014) 105005 [\href{http://arXiv.org/abs/1401.2441}{{\tt
			1401.2441}}].
	
	\bibitem{Alexandrou:2015yba}
	Alexandrou, Constantia and Athenodorou, Andreas and Jansen, Karl, {\it
		{Topological charge using cooling and the gradient flow}},  {\em Phys. Rev. D} {\bf 92}
	(2015) 125014 [\href{http://arXiv.org/abs/1509.04259}{{\tt
			1509.04259}}].
	
	\bibitem{BilsonThompson:2002jk}
	Sundance~O. Bilson-Thompson, Derek~B. Leinweber, and Anthony~G. Williams, {\it
		{Highly improved lattice field strength tensor}},  {\em Annals Phys.} {\bf 304}
	(2003) 1--21 [\href{http://arXiv.org/abs/hep-lat/0203008}{{\tt
			hep-lat/0203008}}].
	
	\bibitem{Bruckmann:2006wf}
	Falk Bruckmann, Christof Gattringer, Ernst-Michael Ilgenfritz, Michael
	Muller-Preussker, Andreas Schafer, and Stefan Solbrig, {\it
		{Quantitative comparison of filtering methods in lattice QCD}},  {\em Eur. Phys. J. A} {\bf 33}
	(2007) 333--338 [\href{http://arXiv.org/abs/hep-lat/0612024}{{\tt
			hep-lat/0612024}}].
		
	\bibitem{Chowdhury:2014mra}
	Abhishek Chowdhury, A.~Harindranath, and Jyotirmoy Maiti, {\it
		{Correlation and localization properties of topological charge density
			and the pseudoscalar glueball mass in su(3) lattice yang-mills theory}},  {\em Phys. Rev. D} {\bf 91}
	(2015) 074507 [\href{http://arXiv.org/abs/1409.6459}{{\tt
			1409.6459}}].	

	
\end{thebibliography}
\end{document}